\documentclass[sigconf]{acmart}
\usepackage{tikz}
\usepackage{amsmath}
\usepackage{amssymb,amsfonts}
\usepackage{algorithmic}
\usepackage{graphicx}
\usepackage{textcomp}
\usepackage{subfigure}
\usepackage[english]{babel}
\usepackage{color}
\usepackage{tabularx}
\usepackage{multirow}
\usepackage{array}
\usepackage{ifpdf}
\usepackage{url}
\usepackage{algorithm}
\usepackage{mdwlist}  %itemize used in the algorithm
\usepackage{caption}
\usepackage{diagbox}

\usepackage{enumitem}
\usepackage{mathrsfs}
\usepackage{pifont}

\newcommand{\ignore}[1]{}
\newcommand{\tabincell}[2]{\begin{tabular}{@{}#1@{}}#2\end{tabular}}
\newtheorem{insight}{Insight}

\AtBeginDocument{%
  \providecommand\BibTeX{{%
    \normalfont B\kern-0.5em{\scshape i\kern-0.25em b}\kern-0.8em\TeX}}}

\acmConference[ICSE 2022]{The 44th International Conference on Software Engineering}{May 21–29, 2022}{Pittsburgh, PA, USA}
\copyrightyear{2022} 
\acmYear{2022} 
\setcopyright{acmcopyright}\acmConference[ICSE '22]{44th International Conference on Software Engineering}{May 21--29, 2022}{Pittsburgh, PA, USA}
\acmBooktitle{44th International Conference on Software Engineering (ICSE '22), May 21--29, 2022, Pittsburgh, PA, USA}
\acmPrice{15.00}
\acmDOI{10.1145/3510003.3510181}
\acmISBN{978-1-4503-9221-1/22/05}

\begin{document}

%%
%% The "title" command has an optional parameter,
%% allowing the author to define a "short title" to be used in page headers.
\title{RoPGen: Towards Robust Code Authorship Attribution via Automatic Coding Style Transformation}

%%
%% The "author" command and its associated commands are used to define
%% the authors and their affiliations.
%% Of note is the shared affiliation of the first two authors, and the
%% "authornote" and "authornotemark" commands
%% used to denote shared contribution to the research.

\author{Zhen Li$^{\ast\dag}$, Guenevere (Qian) Chen$^\ast$, Chen Chen$^\sharp$, Yayi Zou$^\S$, Shouhuai Xu$^\ddag$}
\affiliation{
  \institution{$^\ast$University of Texas at San Antonio, USA}
  \institution{$^\dag$Huazhong University of Science and Technology, China}
  \institution{$^\sharp$Center for Research in Computer Vision, University of Central Florida, USA}
  \institution{$^\S$Northeastern University, China}
  \institution{$^\ddag$University of Colorado Colorado Springs, USA}
  \country{}
}

\email{zh\_li@hust.edu.cn, guenevereqian.chen@utsa.edu, chen.chen@crcv.ucf.edu}
\email{20185258@stu.neu.edu.cn, sxu@uccs.edu}

%%% =========================
\iffalse
\author{Zhen Li}
\affiliation{%
  \institution{
  	%Electrical and Computer Engineering, 
  	University of Texas at San Antonio}
  \city{San Antonio}
  \state{Texas}
  \country{USA}
  \postcode{78249}
  }
\affiliation{%
	\institution{
		%School of Cyber Science and Engineering, 
		Huazhong University of Science and Technology}
	\city{Wuhan}
	\country{China}
	\postcode{430074}
}
\email{zh_li@hust.edu.cn}
%
\author{Guenevere (Qian) Chen}
%\authornote{Corresponding author}
\affiliation{%
  \institution{
  	%Electrical and Computer Engineering, 
  	University of Texas at San Antonio}
\city{San Antonio}
\state{Texas}
\country{USA}
\postcode{78249}}
\email{guenevereqian.chen@utsa.edu}

\author{Chen Chen}
\affiliation{%
	\institution{
		Center for Research in Computer Vision, 
		University of Central Florida}
	\city{Orlando}
	\state{Florida}
	\country{USA}
	\postcode{32816}}
\email{chen.chen@crcv.ucf.edu}

\author{Yayi Zou}
\affiliation{%
	\institution{
		%Software College of Northeastern University, 
		Northeastern University}
	\city{Shenyang}
	%\state{Texas}
	\country{China}
	\postcode{110169}}
\email{20185258@stu.neu.edu.cn}

\author{Shouhuai Xu}
\affiliation{%
	\institution{
		%Department of Computer Science, 
		University of Colorado Colorado Springs}
	\city{Colorado Springs}
	\state{Colorado}
	\country{USA}
	\postcode{80918}}
\email{sxu@uccs.edu}
\fi
%%% ========================

%%
%% The abstract is a short summary of the work to be presented in the
%% article.

\begin{abstract}
Source code authorship attribution is an important problem often encountered in applications such as software forensics, bug fixing, and software quality analysis. 
Recent studies show that current source code authorship attribution methods can be compromised by attackers exploiting adversarial examples and coding style manipulation.
This calls for {\em robust} solutions to the problem of code authorship attribution.
In this paper, we initiate the study on making Deep Learning (DL)-based code authorship attribution robust.  
We propose an innovative framework called {\em \underline{Ro}bust coding style \underline{P}atterns \underline{Gen}eration} (RoPGen),
which essentially learns authors’ unique coding style patterns that are hard for attackers to manipulate or imitate. 
The key idea is to combine {\em data augmentation} and {\em gradient augmentation} at the adversarial training phase. This effectively increases the diversity of training examples, generates meaningful perturbations to gradients of deep neural networks, and learns 
diversified representations of coding styles. 
We evaluate the effectiveness of RoPGen using four datasets of programs written in C, C++, and Java. Experimental results show that RoPGen can significantly improve the robustness of DL-based code authorship attribution, by respectively reducing 22.8\% and 41.0\% of the success rate of targeted and untargeted attacks on average.

\end{abstract}

%%
%% The code below is generated by the tool at http://dl.acm.org/ccs.cfm.
%% Please copy and paste the code instead of the example below.
%%


%%
%% Keywords. The author(s) should pick words that accurately describe
%% the work being presented. Separate the keywords with commas.

%% A "teaser" image appears between the author and affiliation
%% information and the body of the document, and typically spans the
%% page.
%\begin{teaserfigure}
%  \includegraphics[width=\textwidth]{sampleteaser}
%  \caption{Seattle Mariners at Spring Training, 2010.}
%  \Description{Enjoying the baseball game from the third-base
%  seats. Ichiro Suzuki preparing to bat.}
%  \label{fig:teaser}
%\end{teaserfigure}

%%
%% This command processes the author and affiliation and title
%% information and builds the first part of the formatted document.
\maketitle

%\tableofcontents

\section{Introduction}
Software forensics analysis aims to determine whether or not there is software intellectual property infringement or theft associated with some given software code. One useful technique for this purpose is source code authorship attribution \cite{DBLP:journals/spe/BurrowsTZ07,DBLP:conf/gecco/LangeM07}, which aims to identify the author(s) of a given software program \cite{DBLP:journals/csur/KalgutkarKGSM19,DBLP:journals/spe/BurrowsUT14}. This technique has been used for many applications, such as code plagiarism detection,
criminal prosecution (e.g., identifying the author of a piece of malicious code), corporate litigation (e.g., determining whether a piece of code is written by a former employee who violates any non-compete clause of contract),  
bug fixing \cite{DBLP:conf/icse/AnvikHM06,DBLP:conf/icse/RahmanD11}, and software quality analysis \cite{DBLP:conf/icse/ThongtanunamMHI16}.

There are multiple approaches to 
%coping with the problem of 
source code authorship attribution, 
including  statistical analysis \cite{DBLP:journals/compsec/KrsulS97,DBLP:journals/jss/DingS04}, similarity measurement \cite{DBLP:conf/icse/FrantzeskouSGK06,burrows2007source,DBLP:conf/gecco/LangeM07}, and machine learning \cite{DBLP:conf/ccs/AbuhamadAMN18,DBLP:conf/esorics/AlsulamiDHMG17,DBLP:journals/corr/abs-2001-11593,DBLP:journals/fgcs/AbuhamadRAUKN19,DBLP:journals/access/UllahWJAA19,yang2017authorship,DBLP:conf/uss/IslamHLNVYG15,DBLP:journals/spe/BurrowsUT14,pellin2000using}. Recent studies show that current source code authorship identification methods can be compromised by two classes of attacks: the ones exploiting {\em adversarial examples} \cite{DBLP:conf/uss/QuiringMR19,liu2021practical} and the ones exploiting {\em coding style imitation/hiding} \cite{DBLP:journals/popets/SimkoZK18,DBLP:conf/wpes/McKnightG18,DBLP:conf/codaspy/MatyukhinaSPP19}. 
For instance, leveraging adversarial examples \cite{DBLP:conf/uss/IslamHLNVYG15, DBLP:conf/ccs/AbuhamadAMN18} can 
%successfully attack more than 
cause misattribution of more than 99\% software programs in the GoogleCodeJam competition dataset \cite{DBLP:conf/uss/QuiringMR19}; whereas
leveraging the coding style hiding \cite{DBLP:conf/uss/IslamHLNVYG15,DBLP:journals/jss/DingS04,burrows2007source} can cause misattribution %successfully attack all 
of all of the software programs in a GitHub dataset \cite{DBLP:conf/codaspy/MatyukhinaSPP19}. 
The state-of-the-art is that current code authorship attribution methods are vulnerable to these attacks. This calls for research on enhancing the robustness of code authorship attribution methods against attacks.

\vspace{.3em}

\noindent{\bf Our contributions}.
In this paper, we initiate the study on enhancing the robustness of Deep Learning (DL)-based code authorship attribution methods. We choose to focus on this family of methods because they can automatically learn coding style patterns (i.e., avoiding laborious involvement of domain experts) and are very promising for real-world adoption \cite{DBLP:conf/ccs/AbuhamadAMN18,DBLP:conf/esorics/AlsulamiDHMG17,DBLP:journals/corr/abs-2001-11593,DBLP:journals/fgcs/AbuhamadRAUKN19,DBLP:journals/access/UllahWJAA19,yang2017authorship}.  
Effectively, we tackle the following problem: {\em How can we enhance the robustness of DL-based code authorship attribution against attacks?}
For this purpose, we need to address {\em two challenges}.  

The first challenge is to consider more attacks than what have been investigated in the literature; otherwise, the resulting defenses would be specific to the known attacks and will soon become obsolete when new attacks are introduced. This is especially true because the known attacks are geared towards domain expert-defined features \cite{DBLP:conf/codaspy/MatyukhinaSPP19}, which may not be sustainable and would sooner or later need to be replaced by automatic feature learning.
This inspires us to explore new/unknown attacks so that we can design defenses that can enhance robustness against both known and new attacks. For this purpose, we introduce two new attacks which exploit automatic coding style imitation and hiding; these attacks can be applied against both DL-based code authorship attribution and other methods.
The new attacks leverage our systematization of semantics-preserving coding style attributes and transformations, which may be of independent value. 
The attacks are of black-box type because they do not need to know the target code authorship attribution methods; instead, they imitate the target author's coding style or hide the true author's.
%while hiding the true author's.

The second challenge is to design effective defenses against the known and new attacks mentioned above, while accommodating a range of neural network structures (rather than a specific one). To address this challenge, it would be natural to leverage the idea of {\em adversarial training} because it has been widely used in other settings
\cite{DBLP:journals/corr/abs-2102-01356,DBLP:conf/iclr/SchottRBB19,DBLP:conf/icml/MainiWK20}.
However, our experimental results show that such adversarial training approaches applied in these settings \cite{DBLP:journals/corr/abs-2102-01356,DBLP:conf/iclr/SchottRBB19,DBLP:conf/icml/MainiWK20}
%such ideas 
cannot effectively mitigate the known and new attacks mentioned above 
(as what will be described in Table \ref{Table_RQ3_success_rate} of Section \ref{subsec:RQ3}). 
This prompts us to propose an innovative framework, called {\em \underline{Ro}bust coding style \underline{P}atterns \underline{Gen}eration} (RoPGen). 
The key idea is to incorporate {\em data augmentation} and {\em gradient augmentation} to learn robust coding style patterns which are difficult for attackers to manipulate or imitate. 
The role of {\em data augmentation} is to increase the amount and diversity of software programs for training purposes. This is achieved by augmenting programs in two ways: (i) imitating coding styles of other authors; and (ii) perturbing programs' coding styles to a small degree without changing their authorship. 
The role of {\em gradient augmentation} is to learn robust DL models with diversified representations by incurring %meaningful 
perturbations to gradients of deep neural networks.
This is achieved as follows: at each training iteration, we sample multiple sub-networks with a certain fraction of the nodes at each layer of the network; then, we use the sampled sub-networks to construct the network with diversified representations during the weights-sharing training process.
The resulting model learns robust coding style patterns which would be difficult to exploit.
It is worth mentioning that gradient augmentation has been used as a regularization method to alleviate {\em over-fitting} of deep neural networks in image classification \cite{DBLP:conf/nips/YangZ020};
we are the first to use it for {\em robust} authorship attribution.

To evaluate the effectiveness of RoPGen, we use four datasets of programs written in C, C++, and Java, namely GCJ-C++ \cite{DBLP:conf/uss/QuiringMR19}, GitHub-Java \cite{yang2017authorship}, GitHub-C, and GCJ-Java. Among them, GCJ-C++ and GCJ-Java are two sets of programs written by authors who participate in programming competitions for solving a given set of problems;
GitHub-Java and GitHub-C are two sets of real-world programs written by different programmers for varying purposes;
%different software programs by various or same authors).
GitHub-C and GCJ-Java are created for the purpose of the present paper.
Experimental results show that RoPGen can significantly improve the  robustness of DL-based code authorship attribution, respectively reducing the success rate of targeted and untargeted attacks by 22.8\% and 41.0\% on average.
We have made the datasets available at {\tt https://github.com/RoPGen/RoPGen}. We will publish the source code of RoPGen on the same website.

%\smallskip

\vspace{0.3em}
\noindent{\bf Paper organization}.
We discuss the notion of coding styles in Section \ref{sec:Background}, introduce two new attacks in Section \ref{sec:Attack}, describe RoPGen in Section \ref{sec:Defense}, present experimental results in Section \ref{sec:experiments-and-results}, discuss limitations in Section \ref{sec:Limitations} and related prior studies in Section \ref{sec:related_work}, and conclude this paper in Section \ref{sec:Conclusion}.

\section{The Notion of Coding Styles}
\label{sec:Background}
The problem of source code authorship attribution has two variants: single-authorship attribution \cite{DBLP:journals/compsec/KrsulS97,DBLP:journals/jss/DingS04,DBLP:conf/icse/FrantzeskouSGK06,burrows2007source,DBLP:conf/gecco/LangeM07,DBLP:conf/ccs/AbuhamadAMN18,DBLP:conf/esorics/AlsulamiDHMG17,DBLP:journals/corr/abs-2001-11593,DBLP:journals/fgcs/AbuhamadRAUKN19,DBLP:journals/access/UllahWJAA19,yang2017authorship,DBLP:conf/uss/IslamHLNVYG15,DBLP:journals/spe/BurrowsUT14,pellin2000using,abuhamad2021large} vs. multi-authorship attribution \cite{DBLP:journals/popets/DauberCHSWNG19,DBLP:journals/popets/AbuhamadANM20}. Since most studies focus on the former variant while the latter is little understood, we focus on addressing the former variant. 

\vspace{0.3em}

\noindent{\bf Coding style attributes}.
The premise for achieving authorship attribution is that each author has a unique {\em coding style}, which can be defined based on four types of attributes related to programs' layout, lexical, syntactic, and semantic information.
Layout attributes include code indentation, empty lines, brackets, and comments \cite{DBLP:conf/uss/IslamHLNVYG15}.
%which can be directly calculated from the source code. 
Lexical attributes describe tokens (e.g., identifier, keyword, operator, and constant), the average length of variable names, the number of variables, and the number of {\tt for} loop statements \cite{DBLP:conf/uss/IslamHLNVYG15,DBLP:conf/ccs/AbuhamadAMN18}. 
Syntactic attributes describe a program's {\em Abstract Syntax Tree} (AST), including syntactic constructs (e.g., unary and ternary operators) and tree structures (e.g., frequency of adjacent nodes and average depth of AST node types) \cite{DBLP:conf/ccs/AbuhamadAMN18,DBLP:conf/codaspy/MatyukhinaSPP19,DBLP:journals/corr/abs-2001-11593}. 
Semantic attributes describe a program's control flows and data flows (e.g., ``{\tt for}'', `` {\tt while}'', ``{\tt if}, {\tt else if}'', ``{\tt switch}, {\tt case}'', and execution order of statements)
%, and selection of data structures) 
\cite{DBLP:conf/codaspy/MatyukhinaSPP19}.

\begin{table*}[!tbp]
	\vspace{-0.2cm}
	\caption{C, C++, and Java coding style attributes serving as a starting point for robust code authorship attribution
	}
	\vspace{-0.2cm}
	\label{Table_coding_style_types}
	\scriptsize
	%\footnotesize
	\centering
	\begin{tabular}{|c|c|p{.18\textwidth}|p{.415\textwidth}|c|c|c|}
		\hline
		Granularity & Attribute \# & Description & Value & Type & Exhaustive? &  Language\\
		\hline
		{\multirow{7}{*}{Token}} & {\multirow{2}{*}{1}} & {\multirow{2}{*}{Identifier naming method}} & Camel case (e.g., {\tt myCount}), Pascal case (e.g., {\tt MyCount}), words separated by underscores, or identifiers starting with underscores. & {\multirow{2}{*}{Lexical}} & {\multirow{2}{*}{Yes}} & {\multirow{2}{*}{C, C++, Java}}\\
		\cline{2-7}
		& {\multirow{1}{*}{2}} & Usage of temporary variable names & Variable names defined in a compound statement of a function. & {\multirow{1}{*}{Lexical}} & {\multirow{1}{*}{No}} & {\multirow{1}{*}{C, C++, Java}}\\
		\cline{2-7}
		& {\multirow{2}{*}{3}} & Usage of non-temporary local identifier names & Variable names defined in functions but not defined in compound statements, or user-defined function calls. & {\multirow{2}{*}{Lexical}} & {\multirow{2}{*}{No}} & {\multirow{2}{*}{C, C++, Java}} \\
		\cline{2-7}
		& {\multirow{1}{*}{4}} & Usage of global declarations & Global constants declared outside of functions. & {\multirow{1}{*}{Lexical}} & {\multirow{1}{*}{No}} & {\multirow{1}{*}{C, C++}}\\
		\cline{2-7}
		& {\multirow{1}{*}{5}} & Access of array/pointer elements & Use the form of array indexes or pointers, e.g., {\tt arr[i]} and {\tt *(arr+i)}. & {\multirow{1}{*}{Lexical}} & {\multirow{1}{*}{Yes}} & {\multirow{1}{*}{C, C++}}\\
		\hline
		{\multirow{21}{*}{Statement}} & {\multirow{2}{*}{6}} & {\multirow{2}{*}{Location of defining local variables}} & Local variables are defined at the beginning of the variable scope, or each local variable is defined when used for the first time. & {\multirow{2}{*}{Syntactic}} & {\multirow{2}{*}{Yes}} & {\multirow{2}{*}{C, C++, Java}} \\
		\cline{2-7}
		& {\multirow{1}{*}{7}} & Location of initializing local variables & Local variables are initialized and defined in same statements, or in different statements. & {\multirow{1}{*}{Syntactic}} & {\multirow{1}{*}{Yes}} & {\multirow{1}{*}{C, C++, Java}}\\
		\cline{2-7}
		& {\multirow{2}{*}{8}} & Definition (and initialization) of multiple variables with same types & Multiple variables with same types are defined (and initialized) in a statement or in multiple statements. & {\multirow{2}{*}{Syntactic}} & {\multirow{2}{*}{Yes}} & {\multirow{2}{*}{C, C++, Java}}\\
		\cline{2-7}
		& {\multirow{2}{*}{9}} & {\multirow{2}{*}{Variable assignment}} & Multiple variable assignments are in a statement (e.g., {\tt tmp=++i;}) or multiple statements (e.g., {\tt ++i; tmp=i;}). & {\multirow{2}{*}{Syntactic}} & {\multirow{2}{*}{Yes}} & {\multirow{2}{*}{C, C++, Java}}\\
		\cline{2-7}
        & {\multirow{2}{*}{10}} & {\multirow{2}{*}{Increment/decrement operation}} & Use increment (or decrement) operator with different locations, e.g., (i) {\tt i++;} %({\tt i-\,-;}) 
        (ii) {\tt ++i;} %({\tt -\,-i;}) 
        (iii) {\tt i=i+1;}  %({\tt i=i-1;}) 
        (iv) {\tt i+=1;}.% ({\tt i-=1;}). 
        & {\multirow{2}{*}{Syntactic}} & {\multirow{2}{*}{Yes}} & {\multirow{2}{*}{C, C++, Java}}\\
		\cline{2-7}
		& 11 & User-defined data types & Use {\tt typedef} to rename a data type or not. & Syntactic & No & C, C++ \\
		\cline{2-7}
		& 12 & Macros & Use macros to replace constants and expressions or not. & Syntactic & No & C, C++\\
		\cline{2-7}
		& {\multirow{2}{*}{13}} & Included header files or imported classes & Header files included in C/C++ programs and classes imported in Java programs. & {\multirow{2}{*}{Semantic}} & {\multirow{2}{*}{No}} & {\multirow{2}{*}{C, C++, Java}}\\
		\cline{2-7}
		& 14 & Usage of return statements & Use {\tt return 0;} to explicitly return success in {\tt main} function or not. & Semantic & Yes & C, C++\\
		\cline{2-7}
		& 15 & Usage of namespaces & Use namespace {\tt std} or not. & Semantic & Yes & C++\\
		\cline{2-7}
		& 16 & Synchronization with stdio & Enable or remove the synchronization of C++ streams and C streams. & Semantic & Yes & C++\\
		\cline{2-7}
		& 17 & Stream redirection & Use {\tt freopen} to redirect predefined streams to specific files or not. & Semantic & Yes & C, C++\\
		\cline{2-7}

		& {\multirow{2}{*}{18}} & {\multirow{2}{*}{Library function calls}} & C++ library function calls (e.g., {\tt cin}, {\tt cout}) or corresponding C library function calls with the same functionalities (e.g., {\tt scanf}, {\tt printf}). & {\multirow{2}{*}{Semantic}} & {\multirow{2}{*}{Yes}} & {\multirow{2}{*}{C++}}\\
		\cline{2-7}
		& {\multirow{2}{*}{19}} & {\multirow{2}{*}{Memory allocation}} & Static array allocation (e.g., {\tt int arr[100];}) or dynamic memory allocation (e.g., {\tt int *arr=malloc(100*sizeof(int));}). & {\multirow{2}{*}{Semantic}} & {\multirow{2}{*}{Yes}} & {\multirow{2}{*}{C, C++}}\\
		\hline
		{\multirow{4}{*}{Basic block}} & 20 & Loop structures & Use {\tt for} structure or {\tt while} structure. & Semantic & Yes & C, C++, Java\\
		\cline{2-7}
		& 21 & Conditional structures & Use conditional operator, {\tt if-else}, or {\tt switch-case} structure. & Semantic & Yes & C, C++, Java\\
		\cline{2-7}
		& {\multirow{2}{*}{22}} & {\multirow{2}{*}{Compound {\tt if} statements}} & Use a logical operator in an {\tt if} condition (e.g., {\tt if(a \&\& b)}) or use multiple {\tt if} conditions (e.g., {\tt if(a)\{if(b)\{...\}\}}). & {\multirow{2}{*}{Semantic}} & {\multirow{2}{*}{Yes}} & {\multirow{2}{*}{C, C++, Java}}\\
		\hline
		{\multirow{2}{*}{Function}} & {\multirow{2}{*}{23}} & {\multirow{2}{*}{Usage of functions}}
		& The maximum layer number of control statements and loops that are nested within each other, or the number of lines of code in the function.  
		& {\multirow{2}{*}{Semantic}} & {\multirow{2}{*}{No}} & {\multirow{2}{*}{C, C++, Java}}\\
		\hline
	\end{tabular}
	\vspace{-0.2cm}
\end{table*}

Since coding styles and their attributes are related to programming languages, we focus on C, C++, and Java programs because they are widely used, while leaving the treatment of other languages to future studies. Even for these specific programming languages, their coding style attributes are scattered in the literature \cite{DBLP:conf/uss/QuiringMR19,liu2021practical,DBLP:journals/popets/SimkoZK18,DBLP:conf/codaspy/MatyukhinaSPP19}. This prompts us to systematize attributes according to the following observations: 
(i) layout attributes can be easily manipulated by code formatting tools \cite{DBLP:conf/uss/QuiringMR19} (e.g., %TutorialsPoint Formatter \cite{online_java_formatter}, 
Code Beautify \cite{Code_Beautify} and Editor Config \cite{EditorConfig});
%Pretty Printer \cite{Pretty_Printer},  
(ii) those attributes, whose values cannot be automatically modified without changing a program's semantics, would not be exploited by an attacker because they make imitation attacks hard to succeed; and
(iii) those attributes, whose values are rarely used (e.g., making programs unnecessarily complicated), would not be exploited by an imitation attacker.
As highlighted in Table \ref{Table_coding_style_types}, these observations lead to 23 coding style attributes, which span across lexical, syntactic, and semantic information.

\begin{figure*}[!tb]
	\vspace{-0.2cm}
	\centering
	\includegraphics[width=\textwidth]{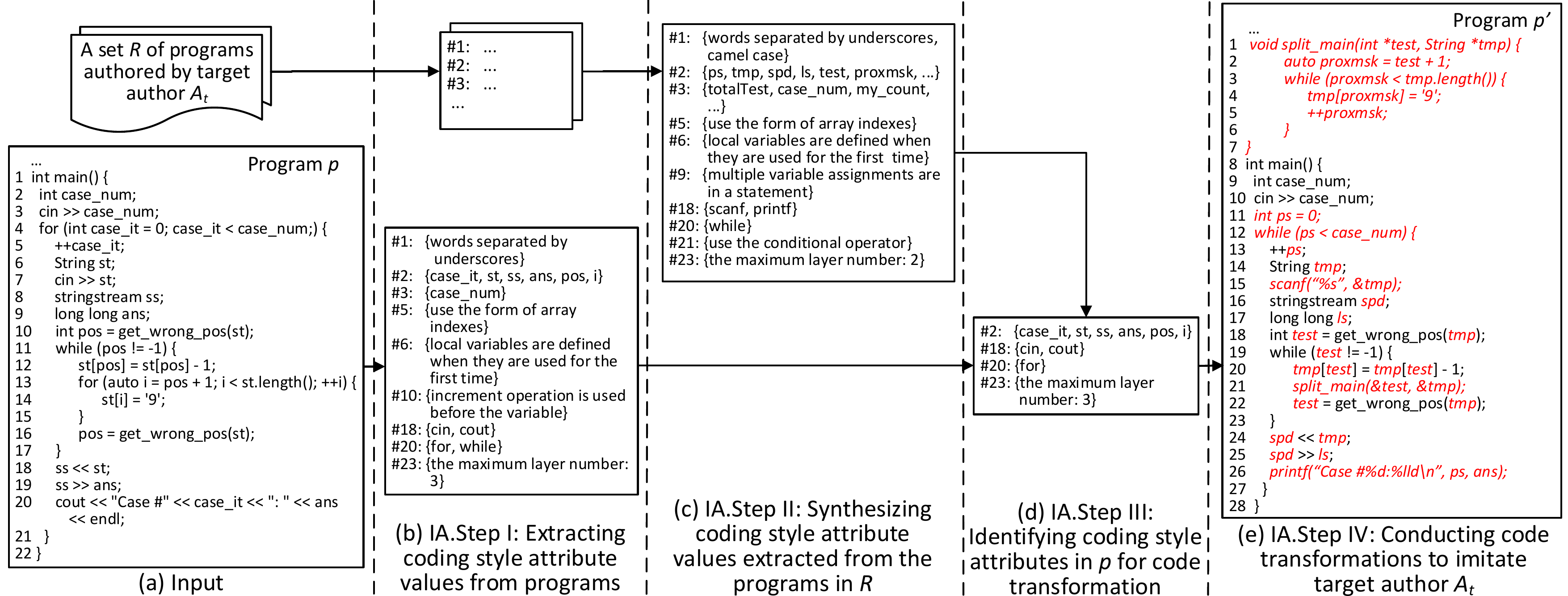}
	\vspace{-0.4cm}
	\caption{An example showing generation of C++ program $p'$ for targeted attack (modified code is highlighted in red and italics)}
	\label{Fig_attack_example}
	\vspace{-0.2cm}
\end{figure*}

\vspace{0.3em}
\noindent{\bf Leveraging coding style attributes as a starting point for robust authorship attribution}. For this purpose, we need to consider two issues. First, we consider {\em granularity} of coding style attributes, namely token vs. statement vs. basic block vs. function. This is important because code transformations on coarse-grained attributes may demand larger degrees of perturbations to programs.
\begin{itemize}
[leftmargin=.32cm,noitemsep,topsep=2pt]
\item {\em Token}-level attributes (\#1-\#5 in Table \ref{Table_coding_style_types}):
They describe the elements in a program's statements: identifier naming method (\#1), usage of temporary variable names (\#2), usage of non-temporary local identifier names (\#3), usage of global declarations (\#4), and access of array/pointer elements (\#5). For instance, attribute \#2 of the program shown in Figure \ref{Fig_attack_example}(a) is described by temporary variable names {\tt case\_it}, {\tt st}, {\tt ss}, {\tt ans}, {\tt pos}, and {\tt i}.

\item {\em Statement}-level attributes (\#6-\#19 in Table \ref{Table_coding_style_types}): They describe  
%Coding styles related to statements involve 
the location of 
%defining or initializing 
defining local variable (\#6), the location of initializing local variables (\#7), the definition (and initialization) of multiple varialbles with same types (\#8), variable assignment (\#9), increment/decrement operation (\#10), user-defined data types (\#11), macros (\#12), included header files or imported classes (\#13), Usage of return statements (\#14), usage of namespaces (\#15), synchronization with stdio (\#16), stream redirection (\#17), library function calls (\#18), and memory allocation (\#19). 
For instance, 
%Take the library functions (i.e., 
attribute \#18 of the program shown in Figure \ref{Fig_attack_example}(a)
%for C++ programs as an example, 
is described by library functions {\tt cin} (Line 7) and {\tt cout} (Line 20).

\item {\em  Basic block}-level attributes (\#20-\#22 in Table \ref{Table_coding_style_types}):
They describe loop structures (\#20), conditional structures (\#21), and compound {\tt if} statements (\#22). For instance, attribute \#20 of the program shown in Figure \ref{Fig_attack_example}(a) is described by two {\tt for} structures (Line 4 and Line 13) and a {\tt while} structure (Line 11).

\item {\em Function}-level attribute (\#23 in Table \ref{Table_coding_style_types}):
At this granularity, coding styles describe %related to function are 
the usage of functions, namely (i) the maximum number of layers of nested compound statements (e.g., control statements and loops) 
%which are nested within each other 
or (ii) the number of lines of code in a function. 
For instance, attribute \#23 of the program shown in Figure \ref{Fig_attack_example}(a) is the 
maximum number of layers of nested compound statements, which is 3 in this case  (i.e., {\tt for}-{\tt while}-{\tt for}).
\end{itemize}

Second, we propose distinguishing those coding style attributes whose domains are exhaustive from those that are not; the term ``exhaustive'' means that an attribute's domain contains few values (e.g., the kinds of loop structures), and a domain is treated as non-exhaustive if its domain contains many values (e.g., the number of possible variable names can be very large). This is important because a non-exhaustive attribute would naturally demand more perturbed examples for adversarial training purposes.
As shown in Table \ref{Table_coding_style_types},
exhaustive attributes include attributes \#1 and \#5 at the token-level granularity, \#6-\#10 and \#14-\#19 at the statement-level granularity, and \#20-\#22 at the basic block-level granularity. For instance, \#20 (i.e., loop structures) has only two values in C, C++, and Java programs: {\tt for} and {\tt while}.
Non-exhaustive attributes include attributes \#2-\#4 at the token level, \#11-\#13 at the statement level, and \#23 at the function level. For instance, \#2 (i.e., usage of temporary variable names) is non-exhaustive because temporary variables can have arbitrary names.
For the program described in Figure \ref{Fig_attack_example}(a), the value of attribute \#2 includes {\tt case\_it}, {\tt st}, {\tt ss}, {\tt ans}, {\tt pos}, and {\tt i}.

\section{Two New Attacks}
\label{sec:Attack}
%In order to make our defense widely applicable, 
We investigate two new attacks against code authorship attribution, 
one is coding style {\em imitation} attack and the other is coding style {\em hiding} attack. 
These attacks are new and can make our defense widely applicable because they are waged {\em automatically} and are waged against both DL-based code authorship attribution and other methods. In contrast, attacks presented in the literature are 
%waged {\em manually}.
manual \cite{DBLP:journals/popets/SimkoZK18}, semi-automatic \cite{DBLP:conf/wpes/McKnightG18}, or automatic but not applicable to DL-based code authorship attribution \cite{DBLP:conf/codaspy/MatyukhinaSPP19}.

Denote by $\mathcal{A}=\{A_1, \ldots, A_\delta\}$ a finite set of authors and by $M$ the code authorship attribution method in question. The attacker has black-box access to $M$, meaning: (i) the attacker can query any program $p$ to $M$ which returns the author of $p$ or $M(p)$; and (ii) how $M$ is obtained is unknown to the attacker. In the threat model, the attacker manipulates $p$ written by $A_s$ (e.g., Alice) into a variant program $p’$ via semantics-preserving code transformations, where $p'\neq p$. The attacker’s goal is:
\begin{itemize}
[leftmargin=.32cm,noitemsep,topsep=2pt]
\item In a {\em targeted} attack with target author $A_t$ (e.g., Bob) where $t \neq s$, the attacker’s goal is to make $M$ misattribute $p’$ to $A_t$, namely $M(p’)=A_t$ while noting that $M$ would correctly attribute $p$ to $A_s$, namely $M(p)=A_s$. That is, the attacker attempts to manipulate a program written by Alice into a semantically-equivalent program which will be misattributed to Bob.
\item In an {\em untargeted} attack, the attacker’s goal is to make $M$ misattribute $p’$ to any other author $A_u$ than $A_s$, namely $M(p’)=A_u$ where $A_u \in \mathcal{A} - \{A_s\}$.
\end{itemize}

\subsection{Automatic Coding Style Imitation Attack}
In this attack, the attacker, $A_s \in \mathcal{A}$ in typical use cases, takes as input: (i) the set $\mathcal{A}$ of authors;
(ii) a program $p$ authored by $A_s$; and (iii) a set $R$ of programs authored by target author $A_t \in \mathcal{A}$ where $t\neq s$. 
The goal of $A_s$ is to {\em automatically} transform program $p$ to program $p'$ such that $p'$ preserves $p$'s functionality and $M$ misattributes $p'$ to $A_t$. The attack proceeds as follows.

\begin{itemize}
[leftmargin=.32cm,noitemsep,topsep=2pt]
\item {\bf IA.Step I: \em Extracting coding style attribute values from program $\boldsymbol{p}$ and the programs in $\boldsymbol{R}$ (authored by target author $\boldsymbol{A_t}$).}
Attacker $A_s$ generates the coding styles of program $p$ and all programs in $R$ 
%Attacker $A_s$ achieves this 
by leveraging the 23 attributes mentioned above (Table \ref{Table_coding_style_types}).
As a running example, Figure \ref{Fig_attack_example}(a) shows $A_s$'s program $p$ and Figure \ref{Fig_attack_example}(b) shows the values of the 9 applicable attributes of $p$. For instance, in order to obtain the value of attribute \#1 (i.e., identifier naming method), $A_s$ can identify all of the user-defined variable and function call names used in $p$ (i.e., {\tt case\_num}, {\tt case\_it}, {\tt st}, {\tt ss}, {\tt ans}, {\tt pos}, {\tt get\_wrong\_pos}, and {\tt i} in this case). Then, $A_s$ can obtain the identifier naming method for each user-defined variable and function call name. Specifically, the value of attribute \#1 corresponding to {\tt case\_num}, {\tt case\_it}, and {\tt get\_wrong\_pos} is ``words separated by underscores''; the other variable and function call names (i.e., {\tt st}, {\tt ss}, {\tt ans}, {\tt pos}, and {\tt i}) cannot be represented by attribute \#1 because these identifiers have no naming rules. Therefore, the value of attribute \#1 of program $p$ is ``words separated by underscores''.

\item {\bf IA.Step II: \em Synthesizing coding style attribute values extracted from the programs in $\boldsymbol{R}$.}
Having extracted attribute values from {\em individual} programs in $R$, we need to synthesize them into a single value for each attribute to obtain target author $A_t$'s coding style.
In the case an attribute is numeric, we propose using the average of an attribute's values (as observed from the programs in $R$) to represent $A_t$'s coding style with respect to the attribute.  In the case an attribute is non-numeric, we propose using the ordered set of an attribute's distinct values 
in the descending order of their frequency to represent $A_t$'s coding style with respect to the attribute. 
As a running example, Figure \ref{Fig_attack_example}(c) illustrates $A_t$'s coding style attributes synthesized from the programs in $R$.
For instance, the synthesized value of numeric attribute \#23 (usage of function) is 2, which is the average of values observed from the programs in $R$. 
Non-numeric attribute \#1 (identifier naming method) takes two distinct values: ``words separated by underscores'' (as observed from most programs in $R$) and ``camel case'' (as observed from the other programs in $R$); the synthesized value of attribute \#1 is the ordered set ``\{words separated by underscores, camel case\}'' as the former has a higher frequency. 

\item {\bf IA.Step III: \em Identifying coding style attributes in $\boldsymbol{p}$ for code transformation.} 
Having obtained attacker $A_s$'s coding style attributes from program $p$ (IA.Step I) and target author $A_t$'s coding style attributes from $R$ (IA.Step II), 
we identify the {\em discrepant} attributes, namely the attributes that take different values with respect to $A_s$ and $A_t$, as candidates for code transformation to make $p$ imitate $A_t$'s coding style. 
For a numeric attribute, discrepancy means that the difference between its value derived from $p$ and its value derived from $R$ is
above a given threshold $\tau$.
For a non-numeric attribute, discrepancy means that its value derived from $p$ is not a subset of its value derived from $R$.
As a running example, Figure \ref{Fig_attack_example} (b) and (c) show that the value of numeric attribute \#23 derived from $p$ is discrepant with the value derived from $R$ 
because their difference, 1, is larger than the threshold $\tau=0$; 
the values of non-numeric attributes \#2, \#18, and \#20 derived from $p$ are discrepant with their counterparts derived from $R$ because the former is not a subset of the latter, respectively.
As shown in Figure \ref{Fig_attack_example} (d), these four discrepant attributes are candidates for code transformations to imitate $A_t$'s coding style.

\item {\bf IA.Step IV: \em Conducting code transformations to imitate target author $A_t$.}
This step is to change the values of the discrepant attributes identified in IA.Step III to imitate target author $A_t$, leading to a transformed (or manipulated) program $p'$ which preserves $p$'s functionality.
We conduct code transformations on individual program files based on {\em srcML} \cite{srcML}, which can preserve program functionalities while supporting multiple programming languages.
As a running example, Figure \ref{Fig_attack_example} (e) shows the manipulated program $p'$ obtained by sequentially transforming the values of attributes \#2, \#18, \#20, and \#23 derived from program $p$, while assuring that each transformation preserves the functionality of the program in question. Take attribute \#23 for example. 
The {\tt main} function (Line 1 in Figure \ref{Fig_attack_example} (a)) is split into two functions {\tt main} (Line 8 in Figure \ref{Fig_attack_example} (e)) and {\tt split\_main} (Line 1 in Figure \ref{Fig_attack_example} (e)).
\end{itemize}

\subsection{Automatic Coding Style Hiding Attack}
In this attack, attacker $A_s \in \mathcal{A}$ takes as input the set $\mathcal{A}$ of authors and a program $p$ authored by $A_s$.
As mentioned above, the goal of $A_s$ is to manipulate program $p$ to another program $p'$, which preserves $p$'s functionality but will not be attributed to $A_s$. 
To achieve this, we propose leveraging the preceding imitation attacks by choosing a target author with the highest misattribution probability. Details follow.

\begin{itemize}
[leftmargin=.32cm,noitemsep,topsep=2pt]
\item {\bf HA.Step I: \em Extracting coding style attribute values from program $\boldsymbol{p}$.}
This is the same as IA.Step I.

\item {\bf HA.Step II: \em Obtaining the coding style of each author $\boldsymbol{A_d}$.} For each $A_d \in \mathcal{A}-\{A_s\}$, we generate $A_d$'s coding style as IA.Step II by treating $A_d$ as the target author.

\item {\bf HA.Step III: \em Identifying the coding style attributes in $\boldsymbol{p}$ for each $\boldsymbol{A_d}$.} For each author $A_d \in \mathcal{A}-\{A_s\}$, we identify the coding style attributes extracted from $p$ that are discrepant with $A_d$’s. This is the same as IA.Step III by treating $A_d$ as the target author. 

\item {\bf HA.Step IV: \em Selecting author $\boldsymbol{A_u}$ for transformation.}
For each $A_d \in \mathcal{A}-\{A_s\}$, we compute the number of lines of code that need to be changed to make $p'$ imitate $A_{d}$'s coding style. 
Changing more lines of code in $p$ (e.g., involving attributes \#11, \#12, and \#13) may make $p$ retain fewer original coding styles and thus make an untargeted attack successful with a higher misattribution probability.
We select author $A_u\in \mathcal{A}-\{A_s\}$  
with the highest misattribution probability as the target author.

\item {\bf HA.Step V: \em Conducting code transformations to imitate author $A_u$.} This is the same as IA.Step IV with target author $A_u$.
\end{itemize}

\section{The RoPGen Framework}
\label{sec:Defense}
In DL-based authorship attribution, the input at the training phase is a set of $\eta$ training programs with labels, denoted by $P=\{p_k,q_k\}_{k=1}^\eta$,
where $p_k$ is a training program and $q_k$ is its label (i.e., author). The output is a DL model $M$. 
Given a finite set of authors $\mathcal{A}=\{A_1, \ldots, A_\delta\}$ and a program $p_k$ authored by $A_s \in \mathcal{A}$, let ${\rm Pr}(M, p_k, A_s)$ denote the probability that $M$ predicts that $p_k$ is authored by $A_s$.
The attacker manipulates $p_k$ to a different program, denoted by $p_k'$. 
As discussed above, an {\em imitation} attacker succeeds when 
${\rm Pr}(M, p_k', A_t)=\max_{1\leq z \leq \delta} \Pr(M, p_k', A_z)$ for a given $t\neq s$;
a {\em hiding} attacker succeeds when ${\rm Pr}(M, p_k', A_s)\neq \max_{1\leq z \leq \delta} \Pr(M, p_k', A_z)$. 

\begin{figure*}[!t]
%	\vspace{-0.2cm}
	\centering
	\includegraphics[width=0.8\textwidth]{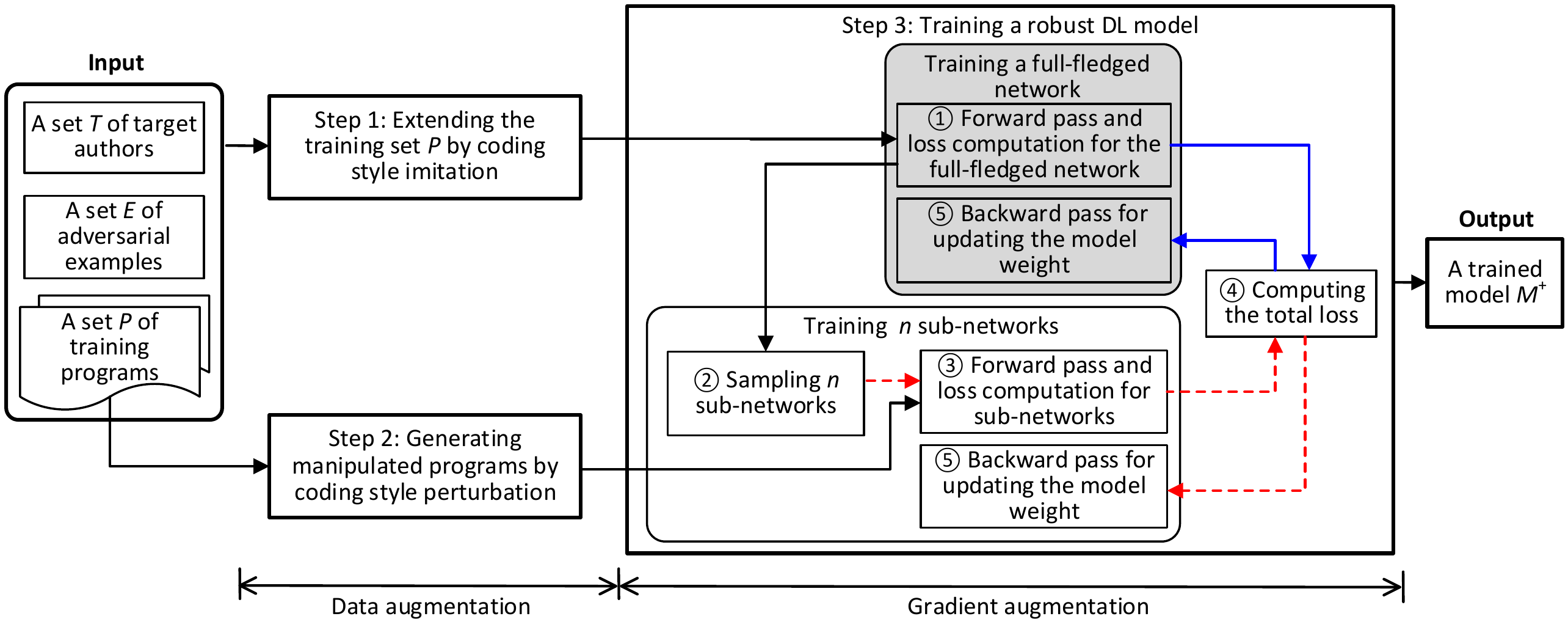}
	\vspace{-0.2cm}
	\caption{\small{The RoPGen framework is an enhanced training model, involving data augmentation (Steps 1 and 2) and gradient augmentation (Step 3).
	Since the data flows share \ding{174} in Step 3, we use solid blue arrows and dotted red arrows to distinguish the training processes of the full-fledged network and sub-networks. 
	The original DL-based training model (baseline) is highlighted with shaded boxes. 
	}}
	\label{Fig_defense_design}
	\vspace{-0.2cm}
\end{figure*}

Figure \ref{Fig_defense_design} highlights the training phase of RoPGen framework, which trains an enhanced model of $M$, denoted by $M^+$. The input to RoPGen includes: (i) a set $P$ of $\eta$ training programs and their labels, (ii) a set $T \subseteq \mathcal{A}$ of target authors, and (iii) a set $E$ of adversarial examples against model $M$.
The basic idea behind RoPGen is to leverage ideas of {\em data augmentation} and {\em gradient augmentation}:

\begin{itemize}
[leftmargin=.32cm,noitemsep,topsep=2pt]
\item {\em  Data augmentation} aims to increase the amount and diversity of training programs. We achieve this via two ideas: (i) imitating coding styles of the other authors,
which is elaborated in Step 1 below; 
(ii) changing programs' coding styles with small perturbations, which is elaborated in Step 2 below. 

\item {\em Gradient augmentation} aims to learn a robust deep neural network with diversified representations by generating meaningful perturbations to gradients. We achieve this by sampling multiple sub-networks, with each involving the first $w_j \times 100\%$ nodes at each layer of the network, where $w_j \in [\alpha, 1]$ and $\alpha$ ($0<\alpha< 1$) is the width lower bound.
This allows a larger sub-network to contain the representation of a smaller sub-network during weights-sharing training, enabling the former to leverage the representations learned by the latter to construct robust networks with diversified representations.
This is elaborated in Step 3 below.

\end{itemize}

\subsection{Step 1: Extending the Training Set by Coding Style Imitation}
Given a set $T$ of target authors,
this step is to extend $P$ by generating programs to imitate the coding styles of the authors in $\mathcal{A}$. 
We first generate a set $P_1$ of programs imitating the coding styles of the authors in $\mathcal{A}$.
Specifically, for each program $p_k \in P$ with label (i.e., authored by) $q_k \in T$, we transform $p_k$ to imitate the coding style of each of the other $\delta-1$ authors in $\mathcal{A}-\{q_k\}$, while preserving $p_k$'s label. 
This essentially repeats the imitation attack described in Section \ref{sec:Attack} for $\delta-1$ times.
Then we obtain the extended set $U=P \cup P_1$ of training programs with labels, which is the input to Step 3 below.

\subsection{Step 2: Generating Manipulated Programs by Coding Style Perturbation}
This step is to generate manipulated programs by coding style perturbation.
We consider two situations.
First, we can generate a set $E$ of adversarial examples against $M$ 
and then obtain a set $U'$ of manipulated programs by leveraging $E$ as follows.
For each adversarial example $e_r\in E$, we obtain a sequence $T_r$ of transformations which led to $e_r$. 
Then, for each program $p_k \in P$, we generate a manipulated program $p_{k,r}$ by conducting the sequence  $T_r$ of transformations. This leads to
$|U'|=|E|\times |P|$ manipulated programs.
Second, if it is not easy to generate adversarial examples, 
%complementary to leveraging adversarial examples, 
we can generate manipulated programs $p_k^1, \ldots, p_k^z$ by perturbing program $p_k$, namely by changing the value of each of the $z$ attributes for each program $p_k \in P$. This leads to a set $U'$ of manipulated programs, where $|U'|=z \times |P|$.
Specifically, we first extract $p_k$'s coding style attributes as in IA.Step I (see Section \ref{sec:Attack}). Corresponding to each attribute $c_j$ ($j=1, \ldots, z$), we generate a manipulated program $p_k^j$ by randomly selecting a value of $c_j$ and changing it to another value, while preserving $p_k$'s label. 
For instance, consider program $p$ in Figure \ref{Fig_attack_example} (a).
For an exhaustive attribute (e.g., attribute \#20), its value (e.g., {\tt while}) can be transformed to another value (e.g., {\tt for}), causing 
the {\tt while} structure (Lines 10 and 11 in Figure \ref{Fig_attack_example} (a)) to be transformed to the {\tt for} structure (i.e., ``{\tt for(pos=get\_wrong\_pos(st); pos!=-1;)\{}'').
For a non-exhaustive attribute (e.g., attribute \#2), its value can be transformed to the value corresponding to another randomly selected author's, causing 
the temporary variable names to become another author's.
Finally, we obtain $U'$ which contains manipulated programs with labels. 

\subsection{Step 3: Training a Robust DL Model $M^+$}
\label{subsec:DL_model}
This step trains a robust model $M^+$ by sampling multiple sub-networks in each training iteration for {\em gradient augmentation} and generating meaningful perturbations to the gradients of the model. 
RoPGen uses the extended training set $U$ as the input to the full-fledged
network and the set $U'$ of manipulated programs 
as the input to the sub-networks.
Denote by $\mathcal{N}$ the deep neural network and $\theta$ its model parameter. Each training iteration has five substeps:

\vspace{0.3em}

%\smallskip
\noindent {\bf Step \ding{172}: Forward pass and loss computation for the full-fledged network.}
We use the extended set $U$ of training programs (obtained in Step 1) as the input to the full-fledged network. 
For each training program with its label $(u,v)\in U$, 
we conduct the forward pass and obtain the predicted value of the full-fledged 
$\mathcal{N}(\theta, u)$. We compute the full-fledged network's loss using the standard
\begin{equation}
\label{equation_loss_full}
L_{std}=l(\mathcal{N}(\theta,u),v)
\end{equation}
and loss function $l$ (e.g., cross entropy).

\vspace{0.3em}

%\smallskip
\noindent {\bf Step \ding{173}: Sampling  $\boldsymbol{n}$ sub-networks.}
We sample $n$ sub-networks $\mathcal{N}_1, \ldots, \mathcal{N}_n$ from the full-fledged network $\mathcal{N}$. To obtain $\mathcal{N}_j$ ($j=1, \ldots, n$), we sample the first $w_j \times 100\%$ nodes in each layer of the full-fledged network. 
The order of nodes at each layer is naturally determined by the full-fledged network (i.e., top-to-bottom in the standard representation of neural networks). We use this order to sample the first $w_j$-fraction of nodes at a layer to obtain a sub-network. 
These sub-networks will be used to learn different representations from manipulated programs and enhance the  robustness of the full-fledged network.

\vspace{0.3em}

%\smallskip
\noindent {\bf Step \ding{174}: Forward pass and loss computation for sub-networks.}
We use $U'$ obtained in Step 2 as the input to each sub-network $\mathcal{N}_j$ because programs in $U'$ are generated with small perturbations and thus suitable for fine-tuning the full-fledged network.
Let $\theta_{w_j}$ be the parameter of the sub-network $\mathcal{N}_j$.
For each program with its label $(u',v') \in U'$, we conduct the forward pass and obtain prediction $\mathcal{N}(\theta_{w_j}, u')$. The loss $L_{subnet}$ of the $n$ sub-networks is
\begin{equation}
\label{equation_loss_sub}
L_{subnet}=\sum_{j=1}^n l(\mathcal{N}(\theta_{w_j}, u'), v'). \\
\end{equation}

\noindent {\bf Step \ding{175}: Computing the total loss.}
The total loss $L_{RoPGen}$ is the sum of the loss of the full-fledged network and the loss of the sub-networks:
\begin{equation}
\label{equation_loss_new}
L_{RoPGen}=L_{std}+L_{subnet}.
\end{equation}
%\smallskip
\noindent {\bf Step \ding{176}: Updating the model weights}.
We conduct the backward pass and leverage the total loss to update model weights, which are shared by the full-fledged network and $n$ sub-networks. 
This allows different parts of the network to learn diverse representations. 

Steps \ding{172} to \ding{176} are iterated until the model converges to $M^+$.

\smallskip
\noindent{\bf Gradient property analysis.} 
To show how Step 3 augments the gradient, it suffices to consider the full-fledged network $\mathcal{N}$ with one layer. Based on Eq. (\ref{equation_loss_full}), the full-fledged network $\mathcal{N}$'s gradient $g_{std}$ is   
\begin{equation}
\label{equation_loss_full_gra}
g_{std}=\frac{\partial l(\mathcal{N}(\theta,u),v)}{\partial \theta}.
\end{equation}
Based on Eq. (\ref{equation_loss_sub}), the $n$ sub-networks' gradient $g_{subnet}$ is 
\begin{equation}
\label{equation_loss_sub_gra}
g_{subnet}=\sum_{j=1}^n \frac{\partial l(\mathcal{N}(\theta_{w_j}, u'), v')}{\partial \theta_{w_j}}.
\end{equation}
Based on Eq. (\ref{equation_loss_new}), Eq. (\ref{equation_loss_full_gra}), and Eq. (\ref{equation_loss_sub_gra}), RoPGen's gradient $g_{RoPGen}$ is
\begin{equation}
\label{equation_loss_new_gra}
g_{RoPGen}=g_{std}+g_{subnet},
\end{equation}
$g_{subnet}$ can be seen as an augmentation to the raw gradient $g_{std}$, explaining the term ``gradient augmentation''.

\section{RoPGen Experiments and Results}
\label{sec:experiments-and-results}

Our experiments aim to answer three Research Questions (RQs):

\begin{itemize}[leftmargin=.5cm,noitemsep,topsep=2pt]
\item {\bf RQ1}: Are the existing DL-based authorship attribution methods robust against the known and new attacks? (Section \ref{subsec:RQ1})
\item {\bf RQ2}: How robust are RoPGen-enabled authorship attribution methods against the known and new attacks? (Section \ref{subsec:RQ2})
\item {\bf RQ3}: Are RoPGen-enabled methods more effective than other adversarial training methods? (Section \ref{subsec:RQ3}) 

\end{itemize}

\subsection{Experimental Setup}

\noindent{\bf Datasets.} Our experiments use four datasets: the first two are used in the literature and the last two are introduced in this paper.

\begin{itemize}[leftmargin=.32cm,noitemsep,topsep=2pt]
\item {\bf \em GCJ-C++ dataset}. 
{\em Google Code Jam} (GCJ) \cite{GCJ} is an annual international programming competition of multiple rounds; each round requires participants to solve some programming challenges.
This dataset is created from GCJ in \cite{DBLP:conf/uss/QuiringMR19} and consists of 1,632 C++ program files from 204 authors. Each author has 8 program files, corresponding to 8 programming challenges, with an average of 74 lines of code per program file.

\item {\bf \em GitHub-Java dataset}. This dataset is created from GitHub in \cite{yang2017authorship} and consists of 2,827 Java program files from 40 authors, with an average of 76 lines of code per program file.

\item {\bf \em GitHub-C dataset}. 
We create this dataset from GitHub, by crawling the C programs of authors who contributed
between 11/2020 and 12/2020.
We filter the repositories that are marked as forks (because they are duplicates) 
and the repositories that simply duplicate the files of others.
We preprocess these files by removing the comments; we then eliminate the resulting files that (i) contain less than 30 lines of code because of their limited functionalities or (ii) overlap more than 60\% of its lines of code with other files.
The resulting dataset has 2,072 C files of 67 authors, with an average of 88 lines of code per file.

\item {\bf \em GCJ-Java dataset}. 
We create this dataset from GCJ between 2015 and 2017.
Since some authors participate in GCJ for multiple years, we merge their files according to their IDs. We select the authors who have written at least 30 Java program files. The dataset has 2,396 Java files of 74 authors, with an average of 139 lines of code per file.
\end{itemize}

\noindent{\bf Evaluation metrics.}
To evaluate effectiveness of code authorship attribution methods, we adopt the widely-used accuracy and attack success rate metrics \cite{DBLP:conf/cvpr/DongFYPSXZ20}. 
Recall that $M$ is a DL-based attribution method, $M^+$ is the RoPGen-enabled version of $M$, and $G$ is an attack method. 
The accuracy of $M$, denoted by $Acc(M)$, is the fraction of the test programs that are correctly labelled by $M$. 
The attack success rate of an imitation attack $G$ against model $M$, denoted by $Asr_{tar}(M,G)$, is the fraction of the manipulated programs that are misattributed to the target author by $M$, among all of the test programs. 
The attack success rate of a hiding attack $G$ against model $M$, denoted by $Asr_{unt}(M,G)$, is the fraction of the manipulated programs that are misattributed to another author by $M$, among the correctly classified test programs.

\vspace{0.3em}
%\smallskip
\noindent{\bf Implementation.}
We choose the following two DL-basd attribution methods reported in  \cite{DBLP:conf/ccs/AbuhamadAMN18,DBLP:journals/corr/abs-2001-11593} because they represent the state-of-the-art and are open-sourced as well as language-agnostic. 

\begin{itemize}[leftmargin=.32cm,noitemsep,topsep=2pt]
\item {\bf \em DL-CAIS}  \cite{DBLP:conf/ccs/AbuhamadAMN18}. This method adopts 
%the {\em Term Frequency-Inverse Document Frequency} (TF-IDF)
lexical features to represent programs, 
leverages recurrent neural network and fully-connected layers to learn representations, and uses random forest to predict authorship. 

\item {\bf \em PbNN} \cite{DBLP:journals/corr/abs-2001-11593}. 
This method adopts code2vec \cite{DBLP:journals/pacmpl/AlonZLY19} to represent programs. It %first 
decomposes a program to multiple paths in its AST, transforms the path-contexts to vectors, 
and uses a fully-connected layer with softmax activation to predict authorship.
\end{itemize}
We use a stratified $\kappa$-fold cross validation, where the dataset is split into $\kappa$-1 subsets for training and the rest for testing.
Following the training strategy of PbNN \cite{DBLP:journals/corr/abs-2001-11593}, we set $\kappa$=10 for the GitHub-C, GCJ-Java, and GitHub-Java datasets.
Following the training strategy of DL-CAIS \cite{DBLP:conf/ccs/AbuhamadAMN18}, we set $\kappa$=8 for the GCJ-C++ dataset.
%which has only 8 program files for each author
This cross validation is repeated $\kappa$ times, where each subset is used for testing the model trained from the other $\kappa$-1 subsets. The evaluation metrics are computed as the average of the $\kappa$ validations.
We use the method reported in 
\cite{DBLP:conf/uss/QuiringMR19} to generate adversarial examples and leverage {\em srcML} \cite{srcML} to generate manipulated  programs and launch coding style imitation/hiding attacks.
We choose {\em srcML} because it can conduct code transformations on an individual program file and can support multiple programming languages.
We conduct experiments on a computer with a NVIDIA
GeForce GTX 3080 GPU and an Intel i9-10900X CPU running at 3.70GHz.

\subsection{Robustness of Existing Methods (RQ1)}
\label{subsec:RQ1}
To determine whether existing authorship attribution methods are robust against the known and new attacks, we attack two DL-based attribution methods (i.e., DL-CAIS \cite{DBLP:conf/ccs/AbuhamadAMN18} and PbNN \cite{DBLP:journals/corr/abs-2001-11593}) on four datasets (i.e., GCJ-C++, GitHub-Java, GitHub-C, and GCJ-Java), corresponding to eight DL models. 

\begin{table}[!tbp]
%	\vspace{-0.2cm}
	\caption{Accuracies of two DL-based attribution methods on four datasets (metrics unit: \%)}
	\vspace{-0.2cm}
	\label{Table_original_accuracy}
	\small
	%\footnotesize
	\centering
	\begin{tabular}{|c|c|c|c|c|}
		\hline
		Method & GCJ-C++ & GitHub-C & GCJ-Java & GitHub-Java \\
		 \hline
		DL-CAIS & 88.2 & 79.9 & 98.5 & 88.4\\
		\hline
		PbNN & 84.8 & 76.7 & 86.2 & 95.4 \\
		\hline
	\end{tabular}
	\vspace{-0.2cm}
\end{table}

Table \ref{Table_original_accuracy} shows 
that DL-CAIS and PbNN on four datasets achieve 88.8\% and 85.8\% accuracies on average. 
For the {\em known} attacks, 
we use the Monte-Carlo tree search to generate adversarial examples \cite{DBLP:conf/uss/QuiringMR19} for each program in the test set of the GCJ-C++ and GitHub-C datasets, since the approach focuses on C/C++ programs. 
To preserve the main coding styles of the original authors, we leverage the notion of {\em $\varphi$-adversary}, which means a program can apply at most $\varphi$ code transformations when generating adversarial examples \cite{DBLP:journals/corr/abs-2002-03043}. 
For the {\em new} attacks, 
we use the automatic coding style imitation and hiding attacks we propose to generate manipulated programs.

\smallskip
\noindent{\bf Robustness against targeted attacks.}
Due to the quadratic number of pairs, we perform targeted attacks on 20 random authors for each dataset and use two program files as the external source (i.e., not part of the training or test set) for extracting each target author's coding style, as per \cite{DBLP:conf/uss/QuiringMR19}. 
For each program authored by these 20 authors in the test set, we respectively take the 19 authors other than the author to whom the program is attributed as the target author. 
For generating adversarial examples, we set $\varphi=3$ (i.e., 3-adversary when generating adversarial examples. We will discuss the impact of different choices of $\varphi$.
Table \ref{Table_RQ1_success_rate} depicts the attack success rates of two DL-based attribution methods on four datasets. 
We observe that the success rate of the targeted attack exploiting adversarial examples 
is 20.3\% lower than that of the targeted attack exploiting coding style imitation
on average. This can be attributed to the fact that adversarial examples obtained by conducting more than three code transformations are not valid attacks with respect to the notion of 3-adversary.
In terms of the time complexity for generating manipulated programs, we consider DL-CAIS on GCJ-C++ dataset as an example. On average, it takes 2,417 seconds to generate an adversarial example of a program; whereas, it only takes 1.5 seconds on average to
generate a manipulated program via the coding style imitation method. 
This large discrepancy can be attributed to the fact that the former method needs to call the attribution model to test candidate examples (possibly multiple rounds in order to generate an adversarial example); whereas, this is not needed in the latter method. For different datasets, the attack success rate of two attribution methods ranges from 9.4\% to 74.6\%, which are related to the number of programs in the dataset and the coding styles of different authors.

\begin{figure}[!htbp]
	%\vspace{-0.4cm}
	\centering
	\includegraphics[width=0.46\textwidth]{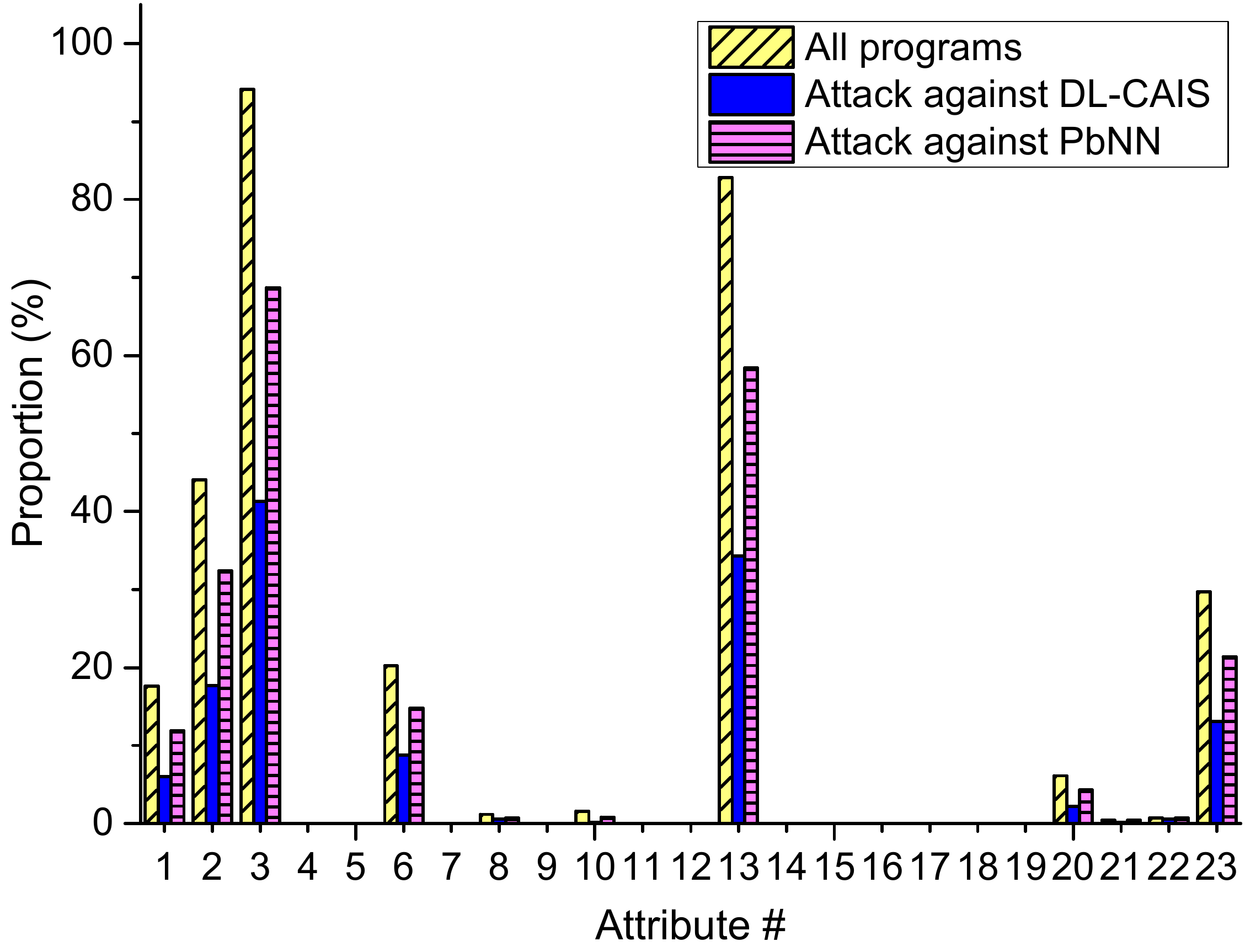}
	\vspace{-0.4cm}
	\caption{Illustrating (i) the proportion of the manipulated programs in the test set involving a coding style attribute's transformation among all manipulated programs in the test set (denoted by ``all programs'') and (ii) the proportion of the manipulated programs that involve a coding style attribute's transformation and can attack successfully in the test set among all manipulated programs in the test set for DL-CAIS and PbNN (denoted by ``attack against DL-CAIS'' and ``attack against PbNN'' respectively).}
	\label{Fig_coding_style_type_distribution}
	\vspace{-0.2cm}
\end{figure}

To see which attributes are changed when generating manipulated programs and the impact of the choice of attributes, let us consider the GCJ-Java dataset.
For each coding style attribute $r$, Figure \ref{Fig_coding_style_type_distribution} illustrates (i) the proportion of the manipulated programs in the test set involving $r$'s transformation among all manipulated programs in the test set and (ii) the proportion of the manipulated programs that involve $r$'s transformation and can attack successfully in the test set among all manipulated programs in the test set for two DL-based attribution methods. We observe that most manipulated programs involve attributes \#1, \#2, \#3, \#6, \#13, and \#23, indicating that these coding style attributes have more significant differences among different authors than other coding style attributes. 
We also observe that the fraction of the manipulated programs that are successful targeted attacks against PbNN is on average 14.4\% higher than that of the successful targeted attacks against DL-CAIS, where manipulations are on attributes \#1, \#2, \#3, \#6, \#13, and \#23.
This indicates that for Java programs,
the path-based representation, which is used by PbNN, can transfer the prediction from one author to another more easily than the token-based representation, which is used by DL-CAIS.

\begin{table}[!tbp]
%	\vspace{-0.4cm}
	\caption{Attack success rates of two DL-based attribution methods, where ``-'' means the method cannot be used on the dataset (metrics unit: \%).}
	\vspace{-0.2cm}
	\label{Table_RQ1_success_rate}
	\small
	\centering
	\begin{tabular}{|c|c|c|c|c|}
		\hline
		Method & GCJ-C++ & GitHub-C & GCJ-Java & GitHub-Java \\
		\hline
		\hline
		\multicolumn{5}{|c|}{Targeted attacks by exploiting adversarial examples ($Asr_{tar}$)}\\
		\hline
		DL-CAIS  & 22.2 & 18.2 & - & - \\
		\hline
		PbNN  & 9.7 & 9.4 & - & -  \\
		\hline
		\hline
		\multicolumn{5}{|c|}{Targeted attacks by coding style imitation ($Asr_{tar}$)}\\
		\hline
		DL-CAIS & 43.9 & 24.3 & 17.7 & 45.1  \\
		\hline
		PbNN & 36.8  & 18.4  & 21.0 & 74.6  \\
		\hline
		\hline
		\multicolumn{5}{|c|}{Untargeted attacks by exploiting adversarial examples ($Asr_{unt}$)}\\
		\hline
		DL-CAIS  & 87.7 & 15.7 & - & - \\
		\hline
		PbNN  & 81.3 & 53.7 & - & - \\
		\hline
		\hline
		\multicolumn{5}{|c|}{Untargeted attacks by coding style hiding ($Asr_{unt}$)}\\
		\hline
		DL-CAIS & 94.8 & 75.0 & 66.3 & 45.0 \\
		\hline
		PbNN & 95.0 & 42.7 & 60.3 & 64.5 \\
		\hline		
	\end{tabular}
%	\vspace{-0.2cm}
\end{table}

\noindent{\bf Robustness against untargeted attacks.} 
We apply the untargeted attack to the correctly classified test programs of authors which are randomly selected in targeted attacks. 
Table \ref{Table_RQ1_success_rate} shows the success rate of untargeted attacks for two DL-based attribution methods on four datasets. We observe that the average success rate of untargeted attacks is 36.8\% higher than that of targeted attacks, which can be attributed to the fact that untargeted attacks, which misattribute program as any author other than the true author, is easier than targeted attacks, which misattribute program to the target author. 
To compare the effectiveness of different methods for coding style hiding attacks, we consider as the baseline a random replacement method, which transforms each coding style attribute value in the program to another random value.  
We choose the random replacement method because it is an intuitive way to make the manipulated program's coding style deviate more from the original author's coding style.

Table \ref{Table_untargeted_attack} summarizes the average results of random replacements five times for each DL model. Our untargeted attack method is significantly better than the random replacement method with 12.7\% higher attack success rate on average. 
This can be explained by the fact that the random replacement method may make the manipulated programs easier to be attributed as the original author because there are some coding style attributes in the program that cannot be automatically transformed. If we do not purposely transform the program's coding style to a target author's, the manipulated program's coding style is more similar to the original author's, causing a failed untargeted attack.

\begin{table}[!tbp]
%	\vspace{-0.2cm}
	\caption{Attack success rates of two methods for coding style hiding attacks (metrics unit: \%)}
	\vspace{-0.2cm}
	\label{Table_untargeted_attack}
	\small
	\centering
	\begin{tabular}{|c|c|c|c|c|}
		\hline
		Method & GCJ-C++ & GitHub-C & GCJ-Java & GitHub-Java \\
		 \hline
		 \hline
		\multicolumn{5}{|c|}{Our untargeted attacks}\\
		\hline
		DL-CAIS & 94.8 & 75.0 & 66.3 & 45.0 \\
		\hline
		PbNN & 95.0 & 42.7 & 60.3 & 64.5 \\
		\hline		
		\hline
		\multicolumn{5}{|c|}{Untargeted attacks by randomly replacement}\\
		\hline		 
		DL-CAIS  & 77.9 & 41.4 & 45.7 & 42.3 \\
		\hline
		PbNN  & 84.0 & 38.0 & 57.1 & 55.7 \\
		\hline
	\end{tabular}
	\vspace{-0.2cm}
\end{table}

To show the impact of $\varphi$ (in $\varphi$-adversary) when generating adversarial examples, we consider DL-CAIS \cite{DBLP:conf/ccs/AbuhamadAMN18} on the GCJ-C++ dataset, while noting that a similar phenomenon is observed for the other DL models. 
%We conduct experiments by using . 
Table \ref{Table_adversary_original} summarizes the attack success rates of DL-CAIS with $\varphi=1, 3, 5$.
We observe that when increasing $\varphi$ from 1 to 5, the attack success rate increases from 5.8\% to 38.8\% for the targeted attack and from 45.2\% to 90.5\% for the untargeted attack. 
This indicates that applying more code transformations can increase the success of imitating or hiding coding styles.

\begin{table}[!tbp]
	%\vspace{-0.2cm}
	\caption{Attack success rates of DL-CAIS method for different $\varphi$-adversaries on the GCJ-C++ dataset (metrics unit: \%) 
	}
	\vspace{-0.2cm}
	\label{Table_adversary_original}
	\small
	\centering
	\begin{tabular}{|c|c|c|c|}
		\hline
		\tabincell{c}{Attack type} & $\varphi=1$ & $\varphi=3$ & $\varphi=5$ \\
		\hline
		Targeted attack & 5.8 & 22.2 & 38.8 \\
		\hline
		Untargeted attack & 45.2 & 87.7 & 90.5 \\
		\hline
	\end{tabular}
	\vspace{-0.2cm}
\end{table}

\begin{insight}
Existing DL-based attribution models are far from robust against the known and new attacks; the success rate of the untargeted attack is much higher than that of the targeted attack because the attacker has more options in the former case.
\end{insight}

\subsection{Robustness of RoPGen (RQ2)}
\label{subsec:RQ2}

To evaluate the effectiveness of RoPGen-enabled authorship attribution methods against known and new attacks, we train eight RoPGen-enabled models involving two DL-based methods
on four datasets. 
We choose the hyperparameters leading to the best accuracy. Take RoPGen-enabled DL-CAIS on the GCJ-C++ dataset as an example. The main hyperparameters are: the batch size is 128, the learning rate is 0.0001, the number of recurrent neural network layers is 3, the width lower bound $\alpha$ is 0.8, 
and the number of sub-networks is 3. 
We set $\varphi=3$ for generating adversarial examples.

Table \ref{Table_RoPGen_accuracy} shows the accuracies of eight RoPGen-enabled models. We observe that the average accuracy of the RoPGen-enabled DL-CAIS models is 2.6\% higher than that of the DL-based models and the average accuracy of the RoPGen-enabled PbNN models is 6.5\% lower than that of the DL-based models, indicating a strong impact of the attribution method.

\begin{table}[!tbp]
%	\vspace{-0.2cm}
	\caption{Accuracies of RoPGen-enabled attribution methods on 4 datasets (metrics unit: \%)}
	\vspace{-0.2cm}
	\label{Table_RoPGen_accuracy}
	\small
	\centering
	\begin{tabular}{|c|c|c|c|c|}
		\hline
		Method & GCJ-C++ & GitHub-C & GCJ-Java & GitHub-Java \\
		 \hline
		DL-CAIS & 92.1 & 84.9 & 98.5 & 90.0\\
		\hline
		PbNN  & 67.6 & 79.7 & 83.6 & 86.1 \\
		\hline
	\end{tabular}
	\vspace{-0.2cm}
\end{table}

Table \ref{Table_RQ2_success_rate} summarizes the attack success rates of RoPGen-enabled methods against attacks. Compared with DL-based attribution methods, RoPGen-enabled methods can reduce 
the success rates of targeted and untargeted attacks (based on exploiting adversarial examples and coding style imitation/hiding) respectively by 22.8\% and 41.0\% on average. 
This means that the RoPGen significantly improves the robustness of DL-based attribution methods against attacks, which can be attributed to the data augmentation and gradient augmentation for learning robust coding style patterns. 
By taking PbNN on the GCJ-C++ dataset as an example, we observe the following. For PbNN, the training phase takes 65.5 seconds; for RoPGen-enabled PbNN, the training phase takes 5,876 seconds (including 5,810.5 seconds incurred by data augmentation and gradient augmentation).
This extra training cost is paid for gaining robustness, while noting that the test cost is almost the same (i.e., 0.010 vs. 0.012 seconds). Since we do not need to train models often, our method is arguably practical.

\begin{table}[!tbp]
%	\vspace{-0.2cm}
	\caption{Attack success rates of RoPGen-enabled attribution methods (metrics unit: \%)
	}
	\vspace{-0.2cm}
	\label{Table_RQ2_success_rate}
	%\small
	\footnotesize
	%\scriptsize
	\centering
	\begin{tabular}{|c|c|c|c|c|}
		\hline
		Method & GCJ-C++ & GitHub-C & GCJ-Java & GitHub-Java \\
		\hline
		\hline
		\multicolumn{5}{|c|}{Targeted attacks by exploiting adversarial examples ($Asr_{tar}$)}\\
		\hline
		RoPGen-enabled DL-CAIS & 19.4 & 3.7 & - & - \\
		\hline
		RoPGen-enabled PbNN & 5.1 & 1.8 & - & - \\
		\hline
		\hline
		\multicolumn{5}{|c|}{Targeted attacks by coding style imitation ($Asr_{tar}$)}\\
		\hline
		RoPGen-enabled DL-CAIS & 3.4 & 1.3 & 0.7 & 0.3 \\
		\hline
		RoPGen-enabled PbNN & 6.3 & 7.2 & 0.6 & 18.0  \\
		\hline
		\hline
		\multicolumn{5}{|c|}{Untargeted attacks by exploiting adversarial examples ($Asr_{unt}$)}\\
		\hline
		RoPGen-enabled DL-CAIS & 58.3 & 9.0 & - & - \\
		\hline
		RoPGen-enabled PbNN & 60.0 & 23.5 & - & - \\
		\hline
		\hline
		\multicolumn{5}{|c|}{Untargeted attacks by coding style hiding ($Asr_{unt}$)}\\
		\hline
		RoPGen-enabled DL-CAIS & 15.0 & 12.4 & 10.9 & 4.2 \\
		\hline
		RoPGen-enabled PbNN & 35.0 & 11.6 & 25.0 & 25.7 \\
		\hline
	\end{tabular}
	\vspace{-0.2cm}
\end{table}

To study the contribution of data augmentation and gradient augmentation to the effectiveness respectively, we conduct the {\em ablation study} to investigate their effects, including three methods. 
The {\em first} method is that we exclude extending the training set by coding style imitation (denoted by ``-CI''), namely the set $P$ of training programs is directly input to the full-fledged network of Step 3.
The {\em second} method is that we exclude the gradient augmentation (denoted by ``-GA''), namely the extended training set $U$ obtained from Step 1 and the set $U'$ of manipulated programs generated from Step 2 together are input to the deep neural network. 
The {\em third} method is that we exclude both coding style perturbation and gradient augmentation from RoPGen (denoted by ``-CP-GA''), namely the extended training set $U$ obtained from Step 1 is input to the deep neural network.

Table \ref{Table_ablation_study} presents the results of applying DL-CAIS \cite{DBLP:conf/ccs/AbuhamadAMN18} to the GCJ-C++ dataset.
We observe that the ``-CI'' method can reduce the success rate of untargeted attacks by exploiting adversarial examples, but are not very effective against targeted attacks by exploiting adversarial examples and coding style imitation and hiding attacks.
The ``-CP-GA'' method can greatly reduce the success rate of coding style imitation and hiding attacks, but are not effective against attacks by exploiting adversarial examples. 
The ``-GA'' method can reduce the success rate of both the coding style imitation and hiding attacks and the attacks by exploiting adversarial examples, but are not as effective as RoPGen.
On average, RoPGen remarkably improves the baseline with a 21.7\% lower success rate of the targeted attack and a 54.6\% lower success rate of the untargeted attack, owing to the incorporation of data augmentation and gradient augmentation.

\begin{table}[!tbp]
%	\vspace{-0.2cm}
	\caption{Ablation analysis results for DL-CAIS on the GCJ-C++ dataset (metrics unit: \%) 
	}
	\vspace{-0.2cm}
	\label{Table_ablation_study}
	\small
	\centering
	\begin{tabular}{|c|c|c||p{.08\textwidth}<{\centering}|p{.08\textwidth}<{\centering}|}
		\hline
	\multirow{2}{*}{Method} & \multicolumn{2}{c||}{Adversarial examples} & \multicolumn{2}{c|}{Coding style imitation/hiding}\\
		\cline{2-5}
		 & $Asr_{tar}$ & $Asr_{unt}$ & $Asr_{tar}$ & $Asr_{unt}$\\
		\hline		
		RoPGen & 19.4 & 58.3 & 3.4 & 15.0 \\
		\hline
		-CI & 27.0 & 61.3 & 25.0 & 65.0\\
		\hline
		-GA & 21.3 & 62.7 & 3.8 & 15.4 \\
		\hline
		-CP-GA & 25.7 & 80.6 & 3.2 & 15.8 \\
		\hline
		Baseline & 22.2 & 87.7 & 43.9 & 94.8 \\
		\hline
	\end{tabular}
	\vspace{-0.2cm}
\end{table}

We evaluate the impact of $\varphi$ in attacks exploiting adversarial examples on the effectiveness of RoPGen-enabled methods. Table \ref{Table_adversary_RoPGen} presents the attack success rate of RoPGen-enabled DL-CAIS on the GCJ-C++ dataset, with $\varphi=1, 3, 5$.
We observe that the attack success rate increases with $\varphi$, exhibiting a similar phenomenon to DL-CAIS; on average, the attack success rate of the RoPGen-enabled DL-CAIS method for targeted and untargeted attacks improves 1.3\% and 23.4\%
with $\varphi$, respectively, compared with the DL-CAIS method (Table \ref{Table_adversary_original}).
This shows the effectiveness of RoPGen-enabled methods against the attacks that exploit adversarial examples.

\begin{table}[!tbp]
	%\vspace{-0.2cm}
	\caption{Attack success rates of RoPGen-enabled DL-CAIS for different $\varphi$ on the GCJ-C++ dataset (metrics unit: \%) 
	}
	\vspace{-0.2cm}
	\label{Table_adversary_RoPGen}
	\small
	\centering
	\begin{tabular}{|c|c|c|c|}
		\hline
		\tabincell{c}{Attack type} & $\varphi=1$ & $\varphi=3$ & $\varphi=5$ \\
		\hline
		Targeted attack & 5.7 & 19.4 & 37.7 \\
		\hline
		Untargeted attack & 28.3 & 58.3 & 66.6 \\
		\hline
	\end{tabular}
%	\vspace{-0.2cm}
\end{table}

\begin{insight}
RoPGen-enabled authorship attribution methods are substantially more robust than the original DL-based methods. In particular, the success rate of targeted and untargeted attacks on RoPGen-enabled methods is respectively reduced by 22.8\% and 41.0\% on average. 
\end{insight}

\subsection{Comparing Adversarial Trainings (RQ3)}
\label{subsec:RQ3} 

To compare the effectiveness of RoPGen-enabled attribution methods with other adversarial training methods, we consider two adversarial training methods from text/source code processing and image classification as baselines, since there have been no defense methods against code authorship attribution attacks so far.
The {\em first} method is basic adversarial training, which is widely used in text processing and source code processing \cite{DBLP:conf/ndss/LiJDLW19,DBLP:conf/aaai/ZhangLLMLJ20}. The basic idea is to generate a set of adversarial examples and adding them to the training set.
%mixing the adversarial examples into the training data to form the augmented training set.
We test two kinds of adversarial examples. One is the adversarial examples generated by \cite{DBLP:conf/uss/QuiringMR19} (denoted by ``Basic-AT-AE''); the other one is the combination of the adversarial examples generated by \cite{DBLP:conf/uss/QuiringMR19} and the programs generated by imitating the coding styles of the authors in $\mathcal{A}$ (denoted by ``Basic-AT-COM'').
The {\em second} method is PGD-AT \cite{DBLP:conf/iclr/MadryMSTV18}, which is a widely-used baseline in image classification. It improves the adversarial robustness by solving the composition of an inner maximization problem and an outer minimization problem. When used to code authorship attribution, PGD-AT has an extremely large search space to search for the coding style transformation with the maximum loss for a program. We use the coding style transformation of a single coding style attribute instead. 

\begin{table}[!tbp]
	\vspace{-0.2cm}
	\caption{Accuracies of DL-CAIS with 4 adversarial training methods on GCJ-C++ and GitHub-C datasets (metrics unit: \%)
	}
	\vspace{-0.4cm}
	\label{Table_RQ3_accuracy}
	\small
	\centering
	\begin{tabular}{|c|c|c|c|}
		\hline
		Method & GCJ-C++ & GitHub-C  \\
		\hline
		None & 88.2 & 79.9 \\
		\hline
		\tabincell{c}{Basic-AT-AE} & 92.6 & 81.5 \\
		\hline
		Basic-AT-COM & 89.2 & 78.2 \\
		\hline
		PGD-AT & 86.2 & 76.1 \\
		\hline
		RoPGen & 92.1 & 84.9 \\
		\hline
	\end{tabular}
	\vspace{-0.2cm}
\end{table}

\begin{table}[!tb]
	%\vspace{-0.4cm}
	\caption{Attack success rates of DL-CAIS with 4 adversarial training methods on the GCJ-C++ and GitHub-C datasets (metrics unit: \%)
	}
	\vspace{-0.2cm}
	\label{Table_RQ3_success_rate}
	\small
	\centering
	\begin{tabular}{|p{.12\textwidth}<{\centering}|p{.12\textwidth}<{\centering}|p{.12\textwidth}<{\centering}|}
		\hline
	    Method & GCJ-C++ & GitHub-C  \\
		\hline
		\hline
		\multicolumn{3}{|c|}{Targeted attacks by exploiting adversarial examples ($Asr_{tar}$)}\\
		\hline
		None & 22.2 & 18.2 \\
		\hline
		\tabincell{c}{Basic-AT-AE}  & 20.4 & 16.5  \\
		\hline
		Basic-AT-COM & 25.4 & 4.2  \\
		\hline
		PGD-AT & 20.6 & 6.9  \\
		\hline
		RoPGen & 19.4  & 3.7  \\
		\hline
		\hline
		\multicolumn{3}{|c|}{Targeted attacks by coding style imitation ($Asr_{tar}$)}\\
		\hline
		None & 43.9 & 24.3  \\
		\hline
		\tabincell{c}{Basic-AT-AE} & 45.7 & 19.9  \\
		\hline
		Basic-AT-COM  & 5.1 & 4.2  \\
		\hline
        PGD-AT  & 24.2 & 6.9  \\
		\hline
		RoPGen  & 3.4 & 1.3  \\
		\hline
		\hline
		\multicolumn{3}{|c|}{Untargeted attacks by exploiting adversarial examples ($Asr_{unt}$)}\\
		\hline
		None & 87.7 & 15.7 \\
		\hline
		\tabincell{c}{Basic-AT-AE}  & 61.4 & 14.8  \\
		\hline
		Basic-AT-COM & 63.5 & 18.5  \\
		\hline
		PGD-AT & 81.7 & 15.0  \\
		\hline
		RoPGen & 58.3 & 9.0  \\
		\hline
		\hline
		\multicolumn{3}{|c|}{Untargeted attacks by coding style hiding ($Asr_{unt}$)}\\
		\hline
		None & 94.8 & 75.0  \\
		\hline
		\tabincell{c}{Basic-AT-AE} & 100.0  & 72.9  \\
		\hline
		Basic-AT-COM  & 15.8 & 27.9  \\
		\hline
        PGD-AT  & 94.2 & 68.0 \\
		\hline
		RoPGen  & 15.0 & 12.4 \\
		\hline
	\end{tabular}
	\vspace{-0.2cm}
\end{table}

Table \ref{Table_RQ3_accuracy} shows the accuracies of DL-CAIS method with four adversarial training methods on the GCJ-C++ and GitHub-C datasets, while noting that PbNN exhibits similar phenomena.
We observe that the accuracies of these adversarial training methods come close to each other, which means these methods have little effect on the accuracy.
Table \ref{Table_RQ3_success_rate} shows the attack success rates of DL-CAIS with four adversarial training methods. 
For {\em Basic-AT-AE} and {\em PGD-AT} methods, the success rate of targeted and untargeted attacks by exploiting adversarial examples is averagely 4.1\% and 8.5\% lower than the original DL-CAIS because a number of manipulated programs with small perturbations are used to improve the model. However, the success rate of coding style imitation/hiding attacks is even a little worse than the original DL-CAIS on some datasets, which means directly extending the training set by programs with small perturbations cannot defend coding style imitation/hiding attacks. 
For {\em Basic-AT-COM} method, the success rate of coding style imitation and hiding attacks is 29.5\% and 63.1\% lower than the original DL-CAIS on average. 
However, the success rate of attacks by exploiting adversarial examples is even a little worse than the original DL-CAIS on some datasets, which means the training set extension with the adversarial examples and the coding styles imitation of other authors cannot defend the attacks by exploiting adversarial examples. 
%Nevertheless, the success rates of attacks by exploiting adversarial examples are still high, leaving a lot of room for improvement.  
Compared with the original DL-CAIS method, RoPGen can reduce the average success rate of targeted and untargeted attacks based on exploiting adversarial examples by 8.7\% and 18.1\% respectively, and reduce the average success rate of targeted and untargeted attacks based on coding style imitation and hiding by 31.8\% and 71.2\% respectively. This attributes to the coding style imitation of other authors, the coding style perturbation, and the gradient augmentation. 

\begin{insight}
Owing to the data augmentation and gradient augmentation, RoPGen substantially outperforms the other adversarial training methods for attacks by both exploiting adversarial examples and coding style imitation/hiding. 
\end{insight}

\section{Limitations}
\label{sec:Limitations}
The present study has several limitations. 
{\bf First}, we focus on improving the robustness of source code authorship attribution methods for a single author owing to its popularity, but the methodology can be adapted to cope with the DL-based multi-authorship attribution methods. Experiments need to be conducted for multi-authorship attribution methods.
{\bf Second}, to evaluate the effectiveness of RoPGen for DL-based attribution methods with different languages, we use two open-source and language-agnostic DL-based attribution methods for evaluation. 
Future studies should investigate other DL-based attribution methods for certain programming languages. 
{\bf Third}, though the RoPGen framework is promising, there is much room for pursuing robust code authorship attribution. Future research should investigate other methods to find the best possible result in defending against attacks.
{\bf Fourth}, for coding style imitation/hiding attacks, we focus on automatic attack methods against code authorship attribution owing to their reproducibility.
It is an interesting future work to investigate whether manual transformation is more powerful than automatic transformation, while noting (i) the manual transformation needs Institutional Review Boards (IRB) approval and (ii) the results would depend on the coding skill of programmers.
{\bf Fifth}, we do not know how to rigorously prove the soundness of various program transformations, but our empirical results provide some hints. 
{\bf Sixth}, it is important to assure the adequacy of threat models.

\section{Related Work}
\label{sec:related_work}
\noindent{\bf Prior studies on {\em non-adversarial} source code authorship attribution.}
Prior studies on non-adversarial authorship attribution can be divided into two categories: {\em single-authorship} attribution \cite{DBLP:journals/compsec/KrsulS97,DBLP:journals/jss/DingS04,DBLP:conf/icse/FrantzeskouSGK06,burrows2007source,DBLP:conf/gecco/LangeM07,DBLP:conf/ccs/AbuhamadAMN18,DBLP:conf/esorics/AlsulamiDHMG17,DBLP:journals/corr/abs-2001-11593,DBLP:journals/fgcs/AbuhamadRAUKN19,DBLP:journals/access/UllahWJAA19,yang2017authorship,DBLP:conf/uss/IslamHLNVYG15,DBLP:journals/corr/abs-2001-11593,DBLP:journals/spe/BurrowsUT14,pellin2000using,abuhamad2021large} vs. {\em multi-authorship} attribution \cite{DBLP:journals/popets/DauberCHSWNG19,DBLP:journals/popets/AbuhamadANM20}. 
There are three approaches to non-adversarial single-authorship attribution  
\cite{DBLP:journals/spe/BurrowsUT14}: (i) the {\em statistical} approach aims to identify important features for discriminant analysis 
\cite{DBLP:journals/compsec/KrsulS97,DBLP:journals/jss/DingS04};
(ii) the {\em similarity} approach uses ranking methods to measure the similarity between 
test examples and candidate examples in the feature space  
\cite{DBLP:conf/icse/FrantzeskouSGK06,burrows2007source,DBLP:conf/gecco/LangeM07};
(iii) the {\em machine learning} approach achieves attribution via random forests \cite{DBLP:conf/uss/IslamHLNVYG15,DBLP:journals/corr/abs-2001-11593}, support vector machines \cite{DBLP:journals/spe/BurrowsUT14,pellin2000using}, and deep neural networks \cite{DBLP:conf/ccs/AbuhamadAMN18,DBLP:conf/esorics/AlsulamiDHMG17,DBLP:journals/corr/abs-2001-11593,DBLP:journals/fgcs/AbuhamadRAUKN19,DBLP:journals/access/UllahWJAA19,yang2017authorship,abuhamad2021large}. 
Whereas, multi-authorship attribution is still largely open \cite{DBLP:journals/popets/DauberCHSWNG19,DBLP:journals/popets/AbuhamadANM20}.
When compared with these studies, we focus on {\em adversarial} single-authorship attribution.

\vspace{0.3em}

%\smallskip
\noindent{\bf Prior studies on {\em adversarial} source code authorship attribution.}
There are two attacks against authorship attribution, which exploit {\em adversarial examples} or {\em coding style imitation/hiding}. 
The former performs functionality-preserving perturbations to a target program to cause misattribution 
\cite{DBLP:conf/uss/QuiringMR19,liu2021practical}.
The latter can be characterized by what the attacker knows
(i.e., black-box \cite{DBLP:journals/popets/SimkoZK18,DBLP:conf/wpes/McKnightG18} vs. white-box \cite{DBLP:conf/codaspy/MatyukhinaSPP19}) 
and what the attacker does
(i.e., manual mimicry attacks \cite{DBLP:journals/popets/SimkoZK18} vs. semi-automatically or automatically leveraging weaknesses of an attribution method \cite{DBLP:conf/wpes/McKnightG18,DBLP:conf/codaspy/MatyukhinaSPP19}).
The most closely related prior study is \cite{DBLP:conf/codaspy/MatyukhinaSPP19}, which presents a white-box attack leveraging human-defined features of the code authorship attribution method.
In contrast, RoPGen deals with black-box attacks which do not know or need such information.
The present study is complementary, or orthogonal, to \cite{DBLP:conf/codaspy/MatyukhinaSPP19} because we focus on coping with black-box attacks against DL-based attribution methods; whereas,
\cite{DBLP:conf/codaspy/MatyukhinaSPP19} %leverages human-defined features and 
cannot deal with DL-based attribution methods because automatically learned features are not human-defined or human-understandable.

\vspace{0.3em}

\noindent{\bf Prior studies on adversarial training.}
From a technical standpoint, RoPGen leverages adversarial training
\cite{DBLP:journals/corr/abs-2102-01356,DBLP:conf/iclr/SchottRBB19,DBLP:conf/icml/MainiWK20}. The basic idea is to augment training data with adversarial examples, analogous to ``vaccination''.
This approach has been extensively investigated in a number of applications, including:
image processing \cite{DBLP:conf/iclr/MadryMSTV18,DBLP:conf/nips/ShafahiNG0DSDTG19,DBLP:conf/iclr/WongRK20,DBLP:conf/icse/GaoSPR20}, neural language processing \cite{wang2019towards,DBLP:conf/ndss/LiJDLW19,DBLP:conf/icml/ZhangAD20}, malware detection \cite{DBLP:conf/uss/Chen0SJ20,DBLP:journals/corr/abs-2005-11671,DBLP:conf/sp/Al-DujailiHHO18,9321695}, and source code processing (e.g., functionality classification, method/variable name prediction, and code summarization) \cite{DBLP:conf/aaai/ZhangLLMLJ20,DBLP:journals/corr/abs-2009-13562,DBLP:conf/icml/BielikV20,DBLP:journals/corr/abs-2002-03043,DBLP:journals/pacmpl/Yefet0Y20,DBLP:conf/iclr/0001C0USLZ21}.
To the best of our knowledge, RoPGen is the first robustness framework for coping with attacks against source code authorship attribution.

\section{Conclusion}
\label{sec:Conclusion}
We presented the RoPGen framework for enhancing robustness of a range of DL-based source code authorship attribution methods. The key idea behind RoPGen is to learn coding style patterns which are hard to manipulate or imitate. This is achieved by leveraging data augmentation and gradient augmentation to train attribution models.
We presented two automatic coding style imitation and hiding attacks. Experimental results show that RoPGen can substantially improve the robustness of DL-based code authorship attribution. The limitations of the present study discussed in Section \ref{sec:Limitations} provide interesting problems for future research.

\begin{acks}
We thank the anonymous reviewers for their constructive comments, which guided us in improving the paper. This work was supported in part by the National Science Foundation under Grants \#1812599, \#2122631, and \#2115134, Army Research Office Grant \#W911NF-17-1-0566, and Colorado State Bill 18-086. 
Zhen Li was supported in part by the National Natural Science Foundation of China under Grant U1936211. 
Any opinions, findings, conclusions or recommendations expressed in this work are those of the authors and do not reflect the views of the funding agencies in any sense.
\end{acks}

\bibliographystyle{ACM-Reference-Format}
\bibliography{bibliography}

\end{document}